\documentstyle[preprint,prb,aps]{revtex} \begin{document} 
\author{L.N.
Bulaevskii$^{1,2}$ and E.M. Chudnovsky$^1$} \address{$^1$ Department of Physics 
and Astronomy, CUNY Lehman College\\ 250 Bedford Park Boulevard West, Bronx, 
NY 10468-1589} 
\address{$^2$ Los Alamos National Laboratory, Los Alamos, NM 87545} 
\title{Ferromagnetic Film on a Superconducting Substrate} 
\maketitle 
\begin{abstract} We study the equilibrium domain structure and magnetic flux around 
a ferromagnetic (FM) film with perpendicular magnetization $M_{0}$ on a 
superconducting (SC) substrate. At $4{\pi}M_{0}<H_{c1}$ the SC is in the Meissner 
state and the equilibrium domain width in the film, $l$, scales as $(l/4{\pi}{\lambda}_{L})=(l_{N}/4{\pi}{\lambda}_{L})^{2/3}$ with the domain width 
on a normal (non-superconducting) substrate, $l_{N}/4\pi\lambda_L \gg 1$.
Here $\lambda_L$ is the London penetration length.  
For $4{\pi}M_{0}>H_{c1}$ and $l_{N}$ in excess of about 35${\lambda}_{L}$, the 
domains are connected by SC vortices.  
We argue that pinning of vortices by magnetic domains in FM/SC multilayers can 
provide high critical currents.
\end{abstract}
\newpage
The interaction between magnetism and superconductivity has been intensively studied 
in 
the past, see, e.g., the review \cite{Bulaevskii}.  
The discovery of high-temperature superconductors and advances in manufacturing of nanoscale multilayered systems have added new dimensions to these studies.  In this 
paper we investigate equilibrium magnetic and superconducting phases in a system consisting of a ferromagnetic (FM) film with perpendicular magnetic anisotropy \cite{remark||} on the surface of a superconductor (SC).  
The interest to such systems is two-fold. Firstly, the needs of magneto-optic 
technology have produced a large variety of magnetic films with perpendicular 
magnetic anisotropy.  Some are synthesized on metallic substrates (e.g., Nb) 
and are well-characterized at room temperature, including their domain structure \cite{Hellman1,Hellman2,Hellman3}. We will show that, as the temperature of such 
a system is lowered, its magnetic state must be affected, in a non-trivial manner, 
by the superconducting transition. Secondly, in a multilayered SC/FM system, the 
domain structure in the FM layers should produce the pinning of superconducting 
vortices which may be significantly stronger than the pinning by magnetic dots \cite{Schuller,Ketterson,Metlushko,Bishop}. 

The system under consideration is shown in Fig.1.  We are assuming no exchange 
of electrons between the FM film and the superconductor. This will be true either 
when the ferromagnet is an insulator or when it is separated from the superconductor 
by a thin insulating buffer layer. Then the FM film and the superconductor are 
coupled only by the magnetic field. (The systems with the exchange of electrons 
between the FM and SC layers have been discussed in Ref.~\onlinecite{bb}.) In 
this case the superconductivity makes a profound effect on the domain structure 
in the FM layer, which must be easy to detect in experiment. The physics behind 
this effect is explained below. Consider a FM film of thickness $d_{M}$, with perpendicular magnetic anisotropy. In the absence of the superconductor adjacent 
to the film, its domain structure is determined by the balance of the energy of 
the magnetic field surrounding the film and the energy of domain walls. The 
positive energy of the magnetic field favors small domains, so that the field 
does not spread too far from the film. On the contrary, the positive energy of 
domain walls  favors less walls, that is, large domains. The minimization of 
the total magnetic energy gives a well-known result \cite{LL} for the equilibrium 
domain width, $l{\propto}\sqrt{{\delta}d_{M}}$, with ${\delta}$ being the domain 
wall thickness. Domains typically observed in magneto-optic films have thickness 
of a few micron. In the presence of a superconductor adjacent to the FM film, 
the balance of the magnetic energy changes drastically. This is because the 
magnetic field must be either expelled from the superconductor due to the 
Meissner effect or it should penetrate into the superconductor in the form 
of vortices. In the first case, the superconductor favors FM domains of width 
below the London penetration depth, ${\lambda}_{L}$. If the room temperature 
domains are significantly greater then ${\lambda}_{L}$, the effect of the SC 
phase transition on the domain structure will be dramatic. As we shall see, 
the new equilibrium will be achieved at 
$l{\propto}({\delta}d_{M}{\lambda}_{L})^{1/3}$. 
Consequently, on lowering the temperature below the SC critical temperature, 
the domains in the FM film can shrink by an appreciable factor. 

We are assuming the stripe domain structure in the FM film. The width of the 
FM domain, $l$, is presumed large compared with the domain wall thickness 
${\delta}$. The latter is the smallest length in our consideration. Two other characteristic length are the thickness of the FM film, $d_M$, and the London 
penetration depth, ${\lambda}_{L}$, of the SC.
In the case of $l<{\lambda}_{L}$ the magnetic flux penetrates into the SC as it 
would penetrate into a normal non-magnetic metal, making superconductivity 
irrelevant. The case of interest is, therefore, $l>{\lambda}_L$. 
We shall begin with the study of the Meissner state, that is, the state where 
equilibrium vortices are absent. 

The free energy functional for the magnetic field, ${\bf B}=[B_x(x,z), 0, B_z(x,z)]$, 
is
\begin{equation}
{\cal F}({\bf B},{\bf M})= {\cal F}_S({\bf B},{\bf M})+ 
{\cal F}_M({\bf B},{\bf M})\;\;, 
\end{equation}
where
\begin{eqnarray}
{\cal F}_S({\bf B}) & = & \frac{1}{8\pi}\int\,dV[ {\lambda}_{L}^{2}
({\nabla}{\times}{\bf B})^{2}+{\bf B}^{2}] \nonumber \\
{\cal F}_M({\bf B},{\bf M}) & = & \int\,dV\left[ \frac{{\bf B}^2}{8\pi}-
{\bf B}\cdot{\bf M}\right]+ {\cal F}_D \;\;. 
\end{eqnarray}
Here ${\cal F}_S$ is the free energy due to the magnetic field in the superconductor, ${\cal F}_M$ is the free energy of the magnetic film and the empty space above the 
film, ${\bf M}(x)$ is the magnetization inside the magnetic film, and ${\cal F}_D$ 
is the energy
of domain walls. At $l>>{\delta}$, a good approximation for $M(x)$ (see Fig.1) 
is the step-like function along the $X$-axis, $M(x)=\pm M_0$ inside the domains. 
Its Fourier expansion is 
\begin{equation}
M(x)=\frac{4M_{0}}{\pi}\sum_{k=0}^{\infty}\frac{\sin[(2k+1)Qx]}{2k+1}\;\;, 
\end{equation} 
where $M_{0}$ is the magnetization and $Q=2{\pi}/l$. For this domain structure 
${\cal F}_D(l)={\sigma}d_M/l$. 
Here ${\sigma}= \sqrt{2}\beta M_0^2\delta/\pi$ is the energy of the unit area 
of the domain wall and ${\beta}M_{0}^{2}$ is the energy density of the 
perpendicular magnetic anisotropy. 

The equilibrium distribution of the magnetic field should be obtained by the 
minimization of ${\cal F}({\bf B},{\bf M})$ at a given configurations of 
magnetic
domains ${\bf M}(x)$. Introducing ${\bf H}={\bf B}-4\pi {\bf M}$, one obtains 
in terms of ${\bf H}$:
\begin{equation}
{\cal F}_M({\bf H},{\bf M})=\int\,dV\left[ \frac{{\bf H}^2}{8\pi}-
2\pi M^2\right]+{\cal F}_D\;\;, \ \ \ {\cal F}_S({\bf H})=
{\cal F}_S({\bf B}={\bf H})\;\;. \label{fun}
\end{equation}
The field ${\bf H}$ is induced by alternating magnetic charges, 
$\nabla\cdot {\bf M}$, on the two surfaces of the magnetic film \cite{sonin}. 
With account of the Maxwell equation,
${\nabla}{\cdot}{\bf B}=0$, it satisfies
\begin{equation}
\nabla\cdot {\bf H}=-4\pi \nabla\cdot {\bf M}=-4\pi [\delta(z)-
\delta(z+d_M)] M(x)\;\;, \ \ \ \nabla\times {\bf H}=0 
\end{equation}
outside the superconductor, that is, inside the magnetic film, in the buffer layer, 
and in the empty space above the film. Here $\delta(z)$ is the delta-function, 
$z=0$ and $z=-d_M$ are coordinates of the film surfaces. Inside the superconductor 
${\bf H}$ satisfies the London equation,
\begin{equation} 
{\nabla}^{2}{\bf H}-{\lambda_L}^{-2}{\bf H} =  0 \;\;, 
\end{equation}
and the boundary condition that ${\bf H}$ is continuous across the interface between
the buffer layer and the superconductor. Equation (6) is valid if the magnetic field changes on the scale greater than the correlation length $\xi$. 
In our case the smallest
relevant scale of spatial variations of the field is the width of FM domains $l$. 
We shall assume that $l>>\xi$, which is relevant to most situations of practical 
interest. 

Solving the above equations we obtain that, due to the domain structure, 
${\bf H}(x,z)$ decays exponentially away from the surfaces. Taking into 
account that $d_M\gg l$ and neglecting exponentially small terms of order 
$\propto \exp[-d_M(4\pi^2l^{-2}+\lambda_L^{-2})^{1/2}]$, we get 
\begin{equation} 
{\bf H}(x,z) =  \sum_{q}{\bf H}_{q}\exp(-q_{z}|\tilde{z}|+iqx)
\end{equation}
for the magnetic field inside the superconductor, the film, and in the 
empty space. 
Here $\tilde{z}$ is the distance along the $z$-axis from the nearest 
film surface. This gives the Fourier components 
\begin{eqnarray} H_{z,q}  & = &
\sum_{k=0}^{\infty}\frac{4M_{0}}{2k+1}\delta[q-Q(2k+1)] \nonumber \\ 
H_{z,-q} & = & -\sum_{k=0}^{\infty}\frac{4M_{0}}{2k+1}\delta[q+Q(2k+1)]\;\;, \end{eqnarray}
and $H_{x,q}=-iq_{z}H_{z,q}/q$ with $q_{z}^{2}=q^{2}+{\lambda}_{L}^{-2}$ inside 
the superconductor and $q_z=q$ elsewhere. Substituting this equilibrium magnetic 
field 
${\bf H}$ at a given $l$ into the free energy functional, Eq.~(\ref{fun}),  
we obtain the following expressions for ${\cal F}_S(l)$ and ${\cal F}_M(l)$ per 
unit area: 
\begin{eqnarray}
&&{\cal F}_S(l)= \frac{4M_0^2}{\pi Q^{2}{\lambda}_{L}} \sum_{k=0}^{\infty}\frac{[1+(2k+1)^{2}Q^{2}{\lambda_L}^{2}]^{1/2}} 
{(2k+1)^{4 }}\;\;, \\
&&{\cal F}_M(l)=3{\cal F}_S(l, \lambda_L^{-1}=0)+{\cal F}_D(l).
\end{eqnarray} 

Above $T_c$ the free energy of the system as a function of $l$ is given by
\begin{equation} 
{\cal F}_N(l)=4{\cal F}_S(l,\lambda_L^{-1}=0)+{\cal F}_D\;\;. 
\end{equation}
The minimization of (11) gives the well known result for the equilibrium width 
of the domains when the superconductor is in the normal state \cite{LL}, 
\begin{equation}
l_{N}=\left[\frac{\sqrt{2}{\pi}}{7\zeta(3)}\right]^{1/2}
(\beta\delta d_M)^{1/2}\;\;.
\end{equation} 

For the superconducting state of the substrate it is convenient to introduce $\bar{l}=l/4{\pi}{\lambda}_{L}$ and ${\bar{l}}_{N}=l_{N}/4{\pi}{\lambda}_L$. 
Then \begin{equation} 
{\cal F}(\bar{l})=\frac{8M_0^2\lambda_L}{\pi}\sum_{k=0}^{\infty} \frac{\bar{l}}{(2k+1)^3}\left\{3+\left[1+\frac{4{\bar{l}}^2}{(2k+1)^2} \right]^{1/2}+\frac{4{\bar{l}}_{N}^2}{{\bar{l}}^{2}}\right\}\;\;.
\label{10}
\end{equation}
The minimization of ${\cal F}$ with respect to ${\bar{l}}$ produces the dependence 
of ${\bar{l}}$ on ${\bar{l}}_N$ shown in Fig.~2. At ${\bar{l}}_{N} \ll 1$ the field penetrates into the SC same way as it penetrates into the normal metal and ${\bar{l}}\approx{\bar{l}}_{N}$ (see insert to Fig.~2). In the opposite case of ${\bar{l}}_{N} \gg 1$, a rather accurate approximation is 
\begin{equation} 
{\bar{l}}\;{\approx}\;{\bar{l}}_{N}^{2/3}\;.
\end{equation} 
We, therefore, conclude that the SC phase transition in the substrate can result 
in a significant shrinkage of the equilibrium domain width in the FM film if the 
substrate is in the Meissner state. 

The Meissner state studied above should always be the case when $4\pi M_0<H_{c1}$. 
At $4{\pi}M_0>H_{c1}$ the equilibrium energy of the Meissner state, ${\cal F}_{M}$, 
should be compared with the energy of the vortex state. Vortex lines connecting 
domains with opposite magnetization can form only if ${\bar{l}}_{N} \gg 1$. The 
energy of the SC in the vortex state may be easily estimated in the limit of 
$M_0\gg 4{\pi}H_{c1}$. In that case the average distance between vortices is 
small compared to their magnetic radius $\lambda_L$ and the average field in 
the SC is close to that in a normal metal. The corresponding total free energy 
of the system is then close to ${\cal F}_{N}$. Some small corrections to that 
energy arise from the vortex line tension and from the repulsion of vortices, 
which is of order $kl(\Phi_0M_0/8\pi\lambda_L^2)\ln(H_{c2}/4\pi M_0)$ ($k{\sim}1$ accounting for the curvature of vortex lines and for their additional energy 
near the FM surface). These corrections are small in the limit of large 
magnetization. Consequently, for the vortex state of the SC substrate, the 
equilibrium domain width in the FM film should be close to that on the normal 
substrate. For the ratio of the equilibrium free energies one obtains 
${\cal F}/{\cal F}_{N}=[6/7{\zeta}(3)]{\bar{l}}_{N}^{1/3}{\approx}0.713{\bar{l}}_{N}^{1/3}$, 
where 
${\cal F}$ is computed at ${\bar{l}}={\bar{l}}_{N}^{2/3}$ and ${\cal F}_{N}$ is 
computed at ${\bar{l}}={\bar{l}}_{N}$. Thus, at $4{\pi}M_0>H_{c1}$ and ${\bar{l}}_{N}{\geq}2.8$ (that is for $l_{N}$ in excess of about 35${\lambda}_{L}$) 
the vortex state should be energetically more favorable than the Meissner state. 

It should be emphasized that all conditions, obtained in this paper, depend on 
temperature through the temperature dependence of ${\lambda}_{L}$ and $l_{N}$. Nevertheless, the factor that determines the maximum shrinkage of equilibrium FM 
domains in the Meissner phase is a universal number, 
max$(l_{N}/l)$=max$({\bar{l}}_{N}^{1/3})$=$7{\zeta}(3)/6{\approx}1.4$. This 
should be
easy to detect in experiment.

The effects described above fall within common experimental range of parameters. 
They will be noticeable if the room-temperature domains in the FM film are wider 
than  $l_{N}{\sim}$0.5 micron for the Nb substrate (${\lambda}_{L}{\sim}40$ nm) or 
wider than $l_{N}{\sim}$1.6 micron for a high-temperature SC 
(${\lambda}_{L}{\sim}130$ nm). Because equilibrium $l_{N}$ depends on the thickness 
of the FM film, $d_{M}$, the above condition on $l_{N}$ translates, through Eq.~(12), 
into the lower bound on $d_{M}$. For, e.g., a TbFe film \cite{Hellman1} (${\beta}{\sim}10^{2}$ and ${\delta}{\sim}$15 nm) the equilibrium domain width 
should decrease below $T_{c}$ in films of thickness greater than 0.3 micron on 
a Nb substrate or in films thicker than 3 micron on a high-$T_{c}$ substrate. 
For TbFe film on a Nb substrate the Meissner state should occur for $l_{N}<$1.4 
micron ($d_{M}<$2.4 micron) and the vortex state should occur at $l_{N}>$1.4 
micron ($d_{M}>$2.4 micron). In the case of a high-$T_{c}$ substrate the 
Meissner state should occur for $l_{N}<$4.5 micron ($d_{M}<$25 micron), 
while at $l_{N}>$4.5 micron ($d_{M}>$25 micron) the vortex state should occur. 

In a real FM film the stripe domains are curved due to thermal fluctuations 
\cite{Abanov} and due to the pinning of domain walls. This, however, should 
not affect our conclusions as long as the corresponding radius of curvature 
of domains is large compared with other characteristic lengths. Since we are 
interested in the equilibrium magnetic structure due to the FM-SC interactions, 
it is important to acknowledge that strong pinning of domain walls by the 
imperfections may prevent the system from reaching that equilibrium. Possible 
ways to study the equilibrium magnetic structures include choosing systems 
with low coercivity (that is, weak pinning of domain walls), or low Curie 
temperature (below the critical temperature of the SC), or rotating the 
system in a slowly decaying magnetic field. It should be also possible to 
extract changes in the magnetic equilibrium from the study of the magnetic 
hysteresis in the FM film above and below $T_{c}$. A large variety of magnetic 
materials should allow experiments in all interesting ranges of temperature 
and coercivity. 

Finally, we would like to comment on the magnetic pinning of vortices in 
SC/FM multilayers. In the past the enhancement of pinning in conventional 
superconductors was done through manufacturing samples with microscopic defects. 
Several new approaches have been developed in recent years. They include 
manufacturing films with micrometer-size holes \cite{Baert} and depositing 
magnetic particles \cite{Otani} or magnetic dots \cite{dots} onto SC films. 
For high temperature superconductors a remarkable effect has been achieved 
in samples with columnar defects produced by heavy ion irradiation \cite{CD}. 
All the above methods are based upon introducing defects that suppress 
superconductivity. The pinning then arises from the tendency of the vortex 
normal core to match with the region where the superconductivity is suppressed. 
The maximal energy per unit length of the vortex, available for such pinning, 
is the condensation energy of Cooper pairs in the volume of the vortex core, $(H_{c}^{2}/8{\pi}){\pi}{\xi}^{2}{\approx} ({\Phi}_{0}/8{\pi}{\lambda}_{L})^{2}$. 
It is easy to see that the pinning of the magnetic flux of vortices in SC/FM 
multilayers can be significantly stronger than the pinning of vortex cores. 
As the problem of statistical mechanics it will be reported elsewhere \cite{APL}. 
Here we shall just estimate the amplitude of the one-vortex pinning potential 
due to magnetic domains. This estimate can be obtained by considering the energy 
of an extra vortex, created by an external magnetic field in the presence of the 
domain structure in FM layers. The upper bound on the magnetic pinning energy is ${\Phi}_{0}M_{0}$ per unit length of the vortex. For $M_{0}$=500 emu/cm$^{3}$ 
it is one hundred times the energy of the  pinning by columnar defects. The 
magnetic pinning of SC vortices will be effective if the field in the FM 
layers does not exceed the coercive field. Above that field the pinning of 
domain walls disappears and FM domains, together with SC vortices, become 
mobile. Thus, strong pinning of vortices favors large coercivity of the FM 
film, the condition opposite to the one required to observe the shrinkage of 
FM domains in the Meissner phase. Some evidence that magnetic dots produce 
more pronounced pinning than non-magnetic dots has been recently demonstrated 
in Nb films \cite{Hoffmann}. Another study \cite{Tejada} indicated that the irreversibility line of an YBCO film is pushed up when a barium ferrite film 
is deposited on the surface of the SC film. Thus, the study of SC/FM multilayers, 
besides being a fascinating "multidimensional" problem, can also be promising in developing SC systems with high critical currents.

This work has been supported by the U.S. Department of Energy through Grant No. 
DE-FG02-93ER45487.

\pagebreak

\pagebreak

FIGURE CAPTIONS\\
{\bf Fig.~1}.\\
FM film with stripe domains on a SC substrate. \\ {\bf Fig.~2}. \\ Dependence of $\bar{l}$ (normalized equilibrium domain width on the SC substrate) on ${\bar{l}}_{N}$ (normalized domain width on the normal substrate).   
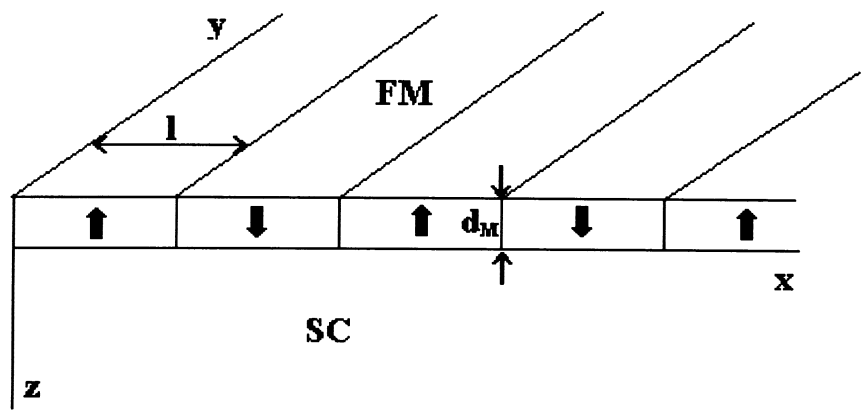
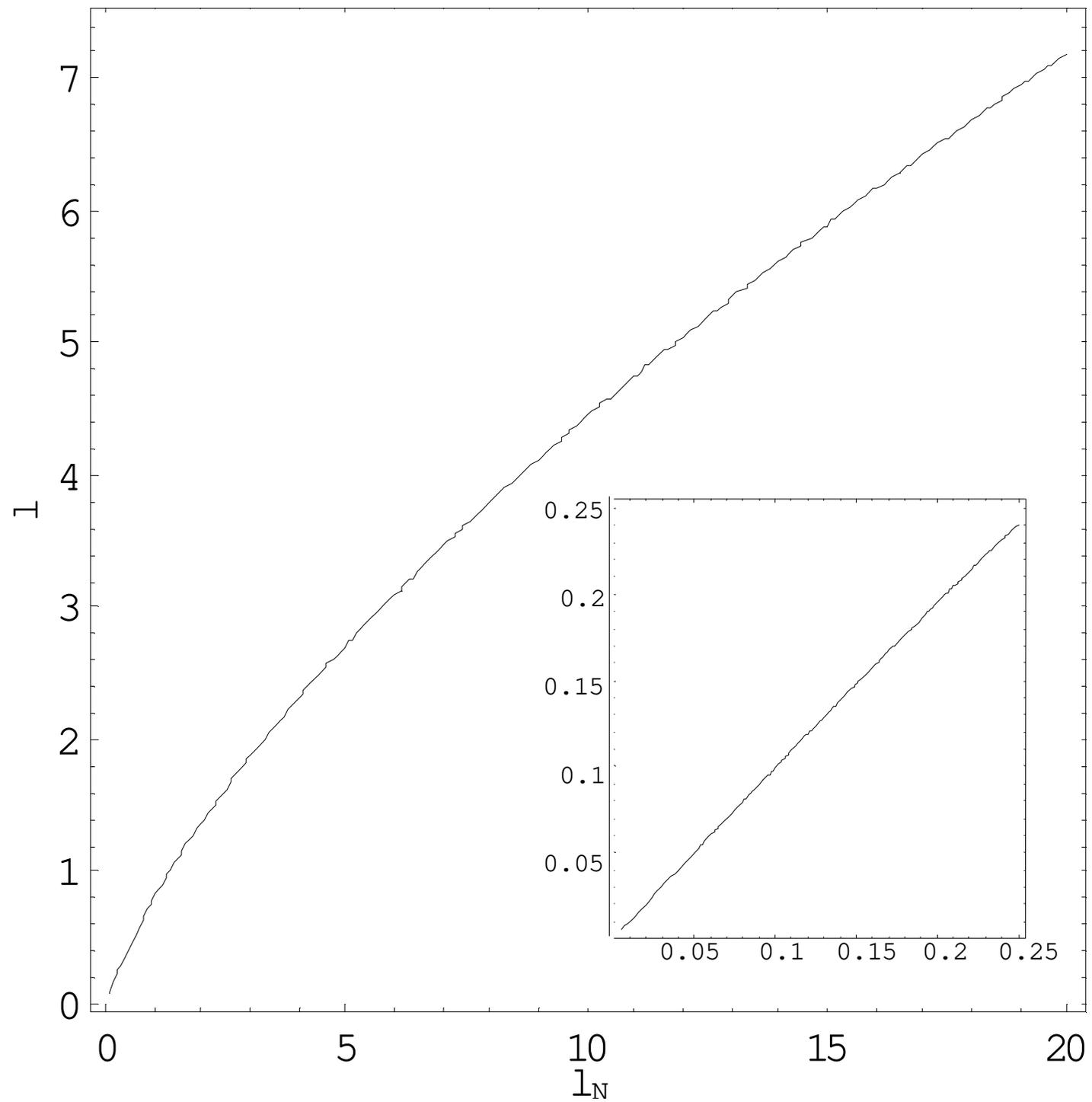
\end{document}